\documentclass[aps,pre,showpacs,twocolumn,groupedaddress,superscriptaddress]{revtex4}
\usepackage{graphicx}%
\usepackage{amsmath}
\usepackage{dcolumn}

\begin{document}

\title{Event--related desynchronization in diffusively coupled oscillator models}

\author{Jane H.~Sheeba}%
\affiliation{Centre for Nonlinear Dynamics, School of Physics,
Bharathidasan University, Tiruchirappalli - 620 024, Tamilnadu, India}

\author{V.~K.~Chandrasekar}%
\affiliation{Centre for Nonlinear Dynamics, School of Physics,
Bharathidasan University, Tiruchirappalli - 620 024, Tamilnadu, India}

\author{M.~Lakshmanan}%
\affiliation{Centre for Nonlinear Dynamics, School of Physics,
Bharathidasan University, Tiruchirappalli - 620 024, Tamilnadu, India}


\begin{abstract}
We seek explanation for the neurophysiological phenomenon of event related desynchronization (ERD) by using models of diffusively coupled nonlinear oscillators. We demonstrate that when the strength of the event is sufficient, ERD is found to emerge and the accomplishment of a behavioral/functional task is determined by the nature of the desynchronized state. We illustrate the phenomenon for the case of limit cycle and chaotic systems. We numerically demonstrate the occurrence of ERD and provide analytical explanation. We also discuss possible applications of the observed phenomenon in real physical systems other than the brain.
\end{abstract}



\maketitle

Brain oscillations at different frequency bands (gamma, theta, beta, alpha, delta waves and so on) are one of the most crucial mechanisms that
control higher level information processing, motor control and large scale integration.
Event--related oscillatory responses of the brain are characterized by means of Event--related
desynchronization/synchronization (ERD/ERS) of neuronal oscillations as observed in
Electroencephalogram (EEG) or Magnetoencephalogram (MEG). ERD causes a relative
decrease in the intensity while ERS causes an increase in the intensity  of a specific frequency band.
Both ERS and ERD are important in deciding the accomplishment of a behavioral/functional
task \cite{Pfurtschelle:99}. Examples include increase in alpha (10-12 Hz) intensity during auditory memory encoding with its suppression during memory retrieval and voluntary movement resulting in a circumscribed desynchronization in the upper alpha and lower theta bands \cite{Pfurtschelle:99}.

\begin{figure}
\begin{center}
\includegraphics[width=8.50cm]{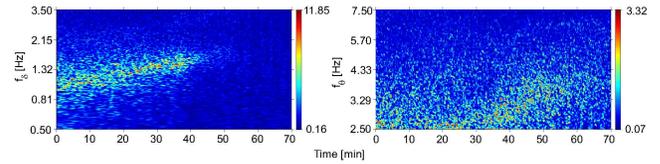}
\caption{(Color online) Experimental observation of dramatic diminution of EEG-delta intensity (left) and emergence of EEG-theta intensity (right) during transition from deep to light anesthesia, the transition time is around 40 mins \cite{Bojan:07}. Note the substantial difference in the intensities of the delta and theta oscillations. Reproduced from Musizza et al. \cite{Bojan:07}, with permission.}
\label{Motiv1}
\end{center}
\end{figure}

Recently, Attention-Deficit/Hyperactivity Disorder (ADHD) is found to be associated with ERD in the alpha
band (8-12 Hz) and ERS in the beta band (15-30 Hz) \cite{Colleen:08}. Dramatic decrease in delta band (1-4 Hz) intensity and an increase in theta (4-8 Hz) intensity is reported to be a signature of transition from deep to light anesthesia \cite{Bojan:07} (see Fig. \ref{Motiv1}). During light anesthesia sensory (external) stimuli from the other parts of the body perturb the synchronized delta oscillations thereby causing desynchronization in the delta band as shown in Fig. \ref{Motiv1} (left) after the transition time. The desynchronized group of neurons contribute to the emergence of theta band oscillations with relatively lesser intensity than those of the delta band oscillations. A model to simulate this experimental observation concluded that some neuronal oscillators desynchronize from the thalamus population (delta band) and synchronize with the cortical population (theta band) for this transition to occur \cite{Jane:08b}. Both ERS/ERD occur due to incoming signals that represent the awaiting task, which may be a motor function, memory tasking or signals from auditory and/or visual cortex that need recognition. Task-specific experimental observations have been made in this direction \cite{Pfurtschelle:99} and task/brain wave - specific detailed models have been proposed for ERD/ERS \cite{Suffczynski:99}; however a more general phenomenological explanation/macroscopic model of the dynamics of ERD is still lacking. We provide here such a dynamical model to this understanding which is very crucial for unfolding the complexity manifold of the brain.

In this Letter, motivated by the phenomenon of ERD, we analyze a system of diffusively coupled population of oscillators subject to an event. Here, by event we mean a dynamic signal from outside the system or may be from a different region in the same system that is characteristically different from the rest of the system. Such a system may be represented by
\begin{eqnarray}
\dot{{\bf X}_j}&=& {\bf F}({\bf X}_j,{\bf \epsilon}_j) + \frac{A}{N}
\sum_{k=1}^{N} ({\bf X}_k-{\bf X}_j) + B {\bf Y}
\label{SL01}
\end{eqnarray}
for $j=1,2, \ldots, N$. Here $N$ represents the size of the system and we take $N=1000$ for all our numerical simulations. The state vector of the $j$th element is represented by ${\bf X}_j$. $A$ is the strength of the coupling between the oscillators. ${\bf Y}$ represents the state vector of the dynamic event which is assumed to evolve according to $\dot{{\bf Y}}={\bf G}({\bf Y},{\bf \epsilon}_e)$ and $B$ is the unidirectional coupling strength of the event. Here ${\bf \epsilon}_e$ represents the parameters of the external system. Typical examples include signals to the brain representing a functional event from the extremities, stimulated microwave current driving a system of spin torque nano--oscillators, polariton condensates in semiconductor micro-cavities that interact both among themselves and with the reservoir, quantum coherence of condensates and so on \cite{Debra:01}. It is well known that for $B=0$, sufficient coupling strength $A>A_c$ causes synchronization in the system leading to ${\bf X}_j={\bf X}$, $j=1, 2, \ldots , N$. In this state all the oscillators behave as one. Now on increasing the strength of the external stimulus (the event) one would expect the synchronized system, which behaves as a single oscillator, to synchronize with the external stimulus. So far, most of the studies have confirmed that, for increasing the stimulus strength $B$, all the oscillators in the system follow certain route/bifurcation patterns to enter into synchronization with the external stimulus \cite{Kuramoto:84}.

However the surprising observation which we report here is that, not all but some of the oscillators in the system separate themselves from the synchronized state and either become desynchronized or synchronized (with an entrainment frequency different from that of the main group) before the whole system of oscillators enter into synchronization with the external stimulus. This intermediate state is crucial for the accomplishment of functional/behavioral tasks in the case of brain, where it becomes necessary that the oscillations at certain frequencies be suppressed (desynchronized) and those at certain other frequencies be enhanced (synchronized) for proper functioning, information processing and cognition via temporal coding. In the case of spin torque nano--oscillators (STNOs), it is generally preferred that all the oscillators are forced to oscillate in synchrony so that the resultant microwave power is larger compared to the power emitted by a single oscillator; hence the intermediate desychronization should be avoided \cite{Pritiraj:05}.
\begin{figure}
\begin{center}
\includegraphics[width=6.5cm]{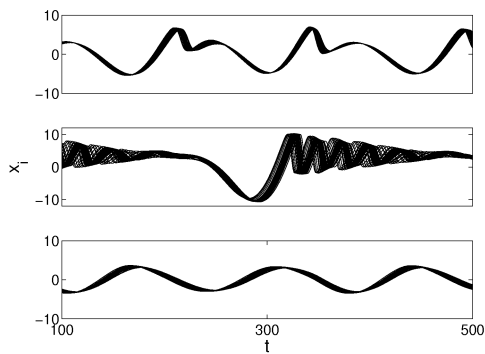}
\caption{The occurrence of ERD in a system of coupled R\"{o}ssler oscillators. Here $A=0.7$, $a=0.1$, $b=0.5$, $c=4$ and $a_e=0.1$, $b_e=0.1$ and $c_e=9$. (top) $B=0.1$, (mid) $B=0.18$ and (bottom) $B=0.23$.}
\label{Ross1}
\end{center}
\end{figure}

We first demonstrate the occurrence of ERD as observed in a system of coupled R\"{o}ssler oscillators described by
\begin{eqnarray}
\dot{x}_j&=&-\omega_jy_j-z_j + B x_e,\nonumber
\end{eqnarray}
\begin{eqnarray}
\dot{y}_j&=&\omega_jx_j+ay_j+ \frac{A}{N}
\sum_{k=1}^{N} (y_k-y_j),\nonumber\\
\dot{z}_j&=&b+z_j(x_j-c),
\label{Ro01}
\end{eqnarray}
which is subject to an external stimulus $x_e$. Here $\omega_j$ are the natural frequencies of the oscillators distributed uniformly between 0 and 1. The system parameters of
the coupled R\"{o}ssler oscillators are chosen so that the oscillators are in the periodic regime, while that of the extermal oscillator are chosen so that it is in the chaotic regime. Coupling with the external oscillator can be in any one of the three components or all the components and the form of the external stimulus is immaterial (periodic, aperiodic, chaotic, etc.) for ERD to be observed.

ERD is depicted in Fig. \ref{Ross1} where the time evolution of the $x$ component of all the oscillators in the system are plotted. We begin with a state of synchronization due to the coupling parameter $A$, when there is no external force, $B=0$. In this state all the oscillators are synchronized completely both in amplitude and phase \cite{ampphase}. Now, for a small strength of the external forcing $B=0.1$, the amplitude synchronization in the system is disturbed (Fig. \ref{Ross1} (top)) while the oscillators still remain phase--locked and hence the time evolution of the state vectors form an evolving band instead of a single evolving line. On increasing $B$ further, desynchronization occurs in both the amplitude and phase, where some oscillators separate themselves from the synchronized group and become desynchronized (Fig. \ref{Ross1} (mid)). Further increase in $B$ brings back synchronization in the system where all the oscillators once again synchronize (Fig. \ref{Ross1} (bottom)). Therefore, the state with which we began was a state of complete synchronization in a specific frequency band that corresponds to a particular functional/behavioral task. In the case of the motivating experiment \cite{Bojan:07}, this is the state of deep an{\ae}sthesia corresponding to synchronization in the $\delta$ frequency band. However, as the strength of the external stimulus (say, the need to code information) increases and when the need to accomplish this task arises (sufficient strength of $B$), the intensity of the previously synchronized frequency band is reduced (which is termed as ERD) by desynchronization of few oscillators from the original frequency. The desynchronized group of oscillators either remains desynchronized (in order to make way for the accomplishment of the incoming task) or synchronizes to a new different frequency (in order to accomplish a different task). During the transition from deep to light an{\ae}sthesia in the experiment \cite{Bojan:07}, the strength of the incoming sensory signals to the brain increases until it becomes sufficiently strong enough to break synchronization in the $\delta$ frequency and engage some of those neuronal oscillators in the $\theta$ band for the accomplishment of coding of information leading to consciousness.
\begin{figure}
\begin{center}
\includegraphics[width=6.5cm]{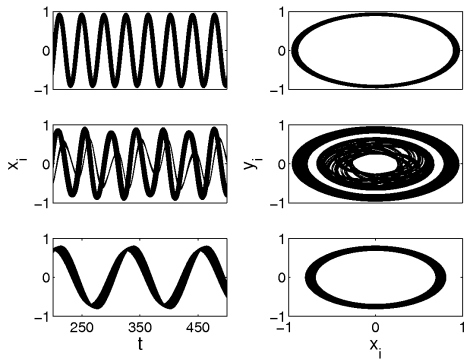}
\caption{The occurrence of ERD in a system of globally coupled SL oscillators. Left panel shows the time evolution of $x$ of all the oscillators. Right panel depicts the corresponding phase portraits on ($x_i,y_i$) plane. Here $A=1.1$, $c=1.5$, $c_{e}=2$, $\omega_e=1.5$ and (top) $B=0.01$, (mid) $B=0.08$, (bottom) $B=0.35$.}
\label{SL1}
\end{center}
\end{figure}

The same phenomenon can also be observed in a system of diffusively coupled Stuart-Landau (SL) oscillators, described by
\begin{eqnarray}
\dot{z}_j=(a+i\omega_j-(1+ic)|z_j|^2)z_j+ \frac{A}{N}\sum_{i=1}^{N} z_i-z_j + B z_{e}.
\label{SLeq01}
\end{eqnarray}
Here $c$ is the nonisochronicity parameter, $a$ is the Hopf bifurcation parameter and  $z_j=x_j+iy_j$ is the complex amplitude of the $j$th oscillator with natural frequency $\omega_j$. $z_{e}$ is another (external) SL oscillator. For more detailed studies about synchronization in populations of SL oscillators one may refer to \cite{Popovych:06,Teramae:04}.

Here also ERD occurs similar to the R\"{o}ssler case, but the separated group of oscillators form a synchronized cluster which can be either periodic or quasi-periodic. The case of quasi-periodic clustering is shown in Fig. \ref{SL1}. The left and the right panels show the time evolution of the state vector $x$ of all the oscillators and the corresponding phase portrait on the ($x_i,y_i$) plane, respectively. The mid panel clearly shows the ERD phenomenon where the separated group oscillates in a quasi-periodic manner. Eventhough the separated group of oscillators are also synchronized within themselves, we will refer to the overall state as being desynchronized since from the point of view of the entire system, this is desynchronization. Desynchronization in the brain is very vital not only for task accomplishment but also in cases of pathologies like Parkinson's disease. A successful method has been proposed in \cite{Popovych:06} to deal with desynchronization in cases of neurological pathologies using delayed feedback. The model discussed in this Letter is also applicable to deal with pathological desynchronization and, in addition, provides a phenomenological explanation for ERD.

\begin{figure}
\begin{center}
\includegraphics[width=4.1cm]{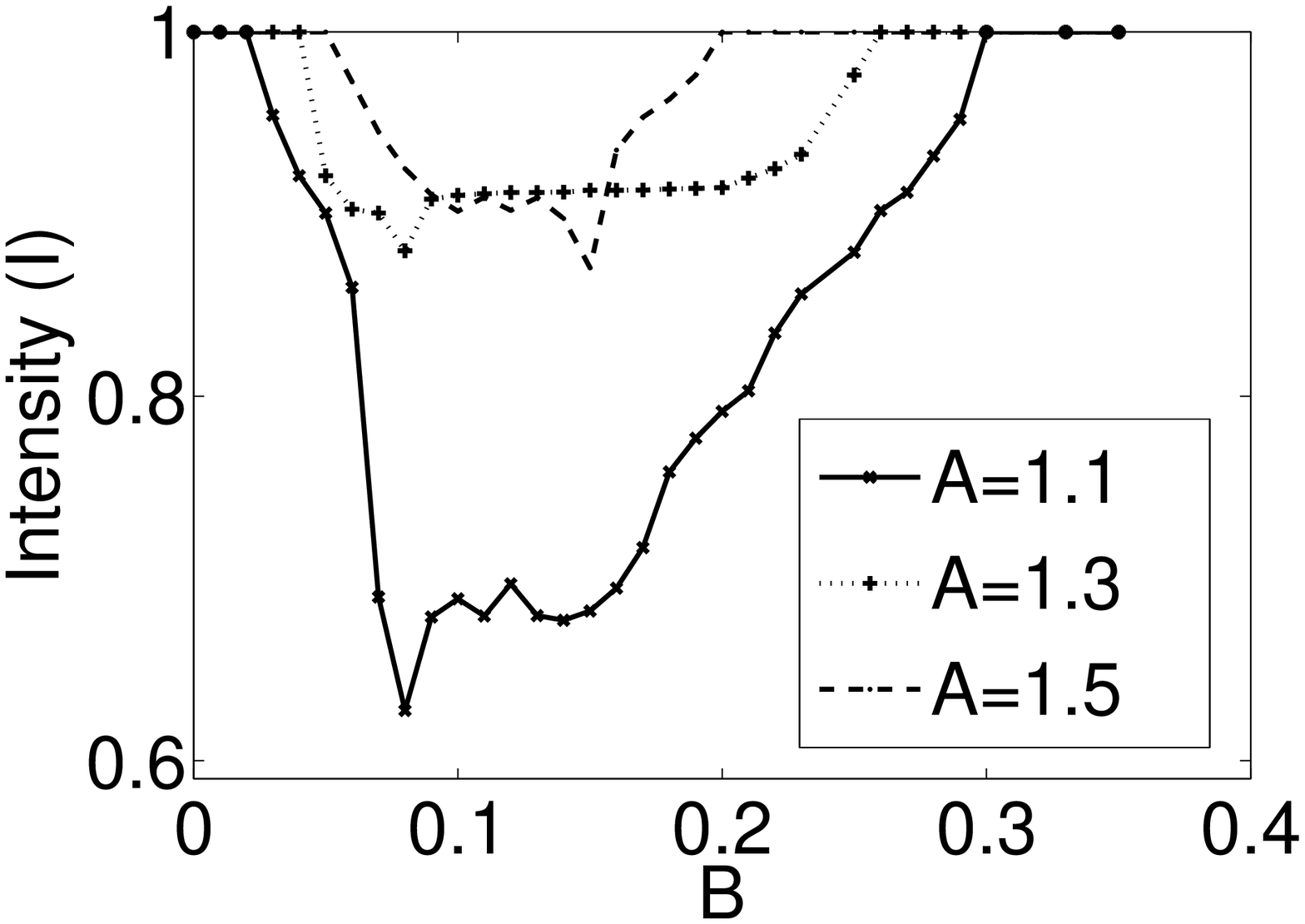}
\includegraphics[width=4.1cm]{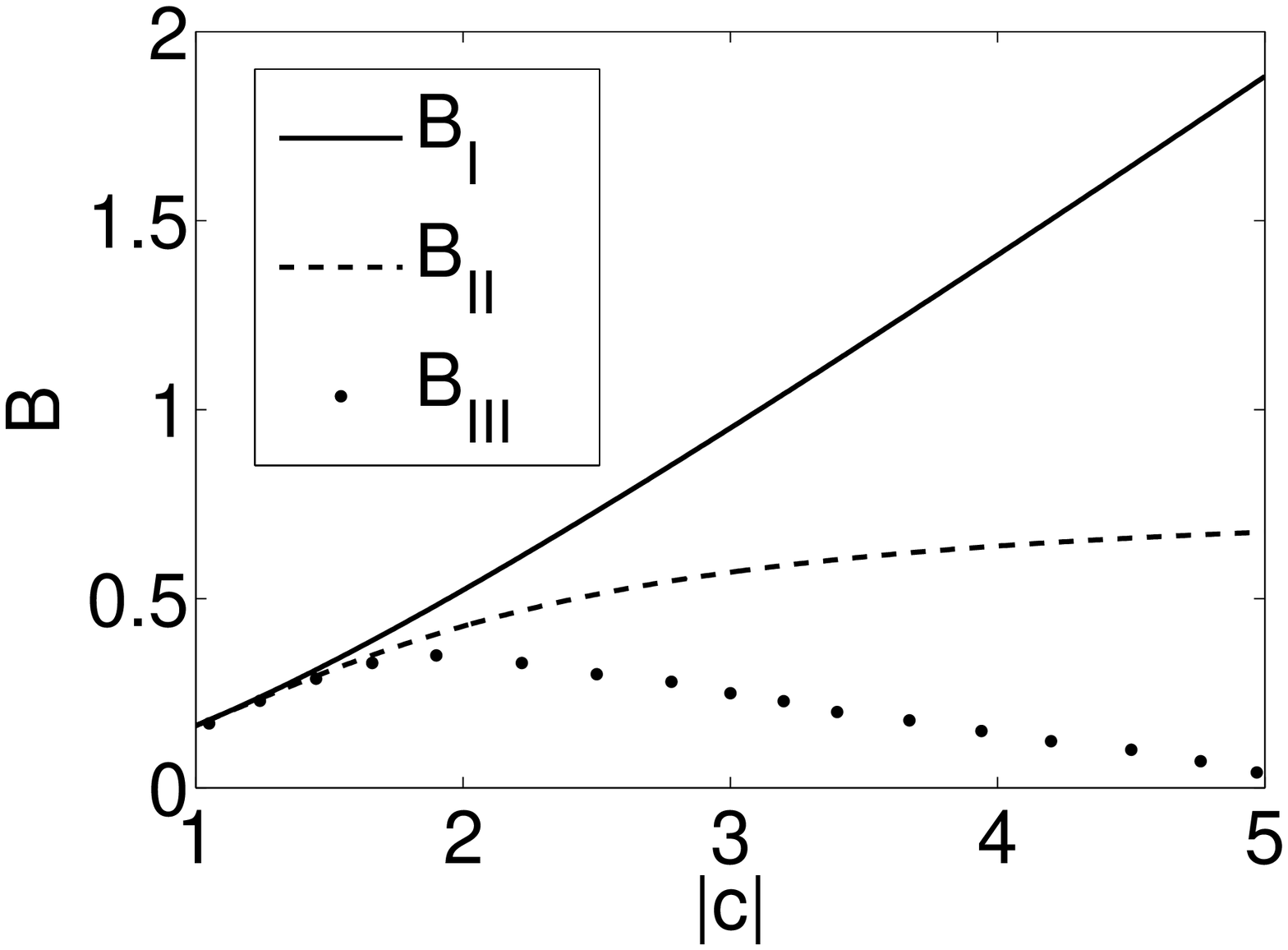}
\caption{(Left) Change in the intensity $I$ due to ERD for the SL system for various values of the coupling strength. (Right) $B-|c|$ phase diagram. $B_I$ and $B_{II}$ are the analytically obtained saddle-node and Hopf bifurcation boundaries respectively. The boundary $B_{III}$ is obtained numerically by solving the evolution equation for $w$ given in the text.}
\label{SLanal1}
\end{center}
\end{figure}

In order to quantify the intensity (which is determined by the number of oscillators that are oscillating in synchrony) of phase synchronization we define the quantity $I=<|\bar{e^{i\theta_j}}|>$, where $\theta_j=\tan^{-1}(y_j/x_j)$ is the phase of the $j$th oscillator. Here the bar represents average over all oscillators in the population and the angle brackets represent time average. The value of $I$ varies between 0 and 1 representing respectively the states of complete desynchronization and complete synchronization. In between these two values, for partial synchronization, $I$ takes a value depending upon the size of the major synchronized cluster; the more oscillators that are oscillating in synchrony the more will be the value of $I$. ERD occurs for those values of the stimulus strength $B$ when the intensity $I$ takes a value less than 1. We use the intensity of \emph{phase} synchronization to characterize the occurrence of ERD because, eventhough amplitude synchronization is disturbed for small values of $B$, ERD occurs at those (sufficient) values of $B$ when both amplitude and phase synchronizations are lost. Therefore monitoring the intensity of phase synchronization will facilitate monitoring the occurrence of ERD. Fig. \ref{SLanal1} (Left)) shows that there exists a critical strength of the external stimulus for ERD to occur. It may also be noted that the ERD occurs for a finite window of the stimulus strength which depends upon the coupling strength $A$. When the coupling strength is just enough to synchronize the system, say $A=1.1$ as shown in Fig. \ref{SLanal1} (Left)), ERD is observed for a wider range of the stimulus strength as compared to stronger coupling strengths, say $A=1.3$ and $A=1.5$.

In both the above systems, numerical simulations show that the separated group comprises of few oscillators (say $N_2$) compared to the size of the major synchronized group (say $N_1$). Therefore the size ratio of the two cluster state is $r:1-r(=N_1/N:N_2/N)$ and $r>>0$ ($\sim 1$) in the desynchronized state. For the case of SL oscillators, the dynamics of the large group is given by
$\dot{u} \approx ((a+i\omega)-(1+ic)|u|^2)u+B e^{i\omega_e t}$. In order to analytically explain the occurrence of desynchronization, we assume that the major synchronized group is completely synchronized with the external force $z_{e}=e^{i\omega_e t}$, while the small group is not. This assumption yields $\omega=\omega_e+c$ and $a=1-B$.
Now, the dynamics of the small group can be approximated in the form
$\dot{w}= ((a+\hat{A}+ic)-(1+ic)|w|^2)w+\hat{A}+B$,
where $w=ve^{-i\omega_e t}$ and $\hat{A}=rA$. The occurrence of desynchronization can be understood from the fixed points of this equation. One of the fixed points, $w_1=1$, is stable which corresponds to complete synchronization of the population, that is $u=v$ \cite{Daido:07}.  The other two fixed points ($w_2$ and $w_3$) that determine the stability of the desynchronized state exist for $B<B_I=(1+c^2)(\sqrt{1+c^2}-1)/(2c^2)-\hat{A}$, where $B_I$ is a saddle-node bifurcation point.

From the linear stability analysis we find that the determinant of the Jacobian matrix is always negative (positive) for $w_2$ ($w_3$). Therefore the fixed point $w_2$ turns out to be a saddle. The other fixed point $w_3$ is either an unstable node or a focus for $|c|<1$. When $|c|>1$ and $B>B_{II}$, the fixed point $w_3$ is either a stable node or a focus; here $B_{II}$ is a Hopf bifurcation point given by $B_{II}=(1+c^2)/(\sqrt{4+(1+c^2)^2}+2)-\hat{A}$ which is determined from the condition $\mbox{tr}(J_{w_3})=0$. The desynchronized solution does not occur for a value of $B$, say $B<B_{III}$, and only synchronized solutions are stable. At this point, both the saddle and the Hopf bifurcation points merge and disappear. Fig. \ref{SLanal1} (Right) gives the bifurcation boundaries where the $B_{III}$ boundary is obtained numerically by solving the evolution equation for $w$. For given values of parameters, desynchronized solutions exist in the region between $B_I$ and $B_{III}$. This is in agreement with our numerical observations. Similar analytical treatment for the case of R\"{o}ssler system can be provided using the idea of co-evolving ampltudes and phases \cite{Aoki:09} of the synchronized and desynchronized groups.

To conclude, we demonstrate that systems of diffusively coupled nonlinear oscillators subject to external stimulus typically act as conceptual models that explain ERD. We explain the dramatic diminution of $\delta$ wave intensity during the transition from deep to light an{\ae}sthesia in a physiological experiment using this model. Although we have used systems of coupled R\"{o}ssler and SL oscillators for illustration, this phenomenon is common in all kinds of diffusively coupled models. Models discussed in this Letter can account for the desynchronization phenomenon due to external stimulus (while all the oscillators in the system are synchronized due to coupling strength in the absence of external stimulus) in systems of STNOs, polariton condensates, neuronal networks, Bose--Einstein condensates, Josephson junction arrays and lasers and so on.

JHS and VKC wish to thank Aneta Stefanovska and Peter V. E. McClintock for useful discussions. The work is supported by a Department of Science and Technology (DST), Government of India -- Ramanna Fellowship program and also by a DST -- IRHPA research project.


\begin{thebibliography}{10}
\bibitem{ampphase}
Here, by amplitude and phase we mean the quantities $\sqrt{x^2+y^2}$ and $\tan^{-1}(\frac{y}{x})$, respectively.
\end{thebibliography}
\end{document}